\documentclass{llncs}
\usepackage{amsfonts}
\usepackage{caption}
\usepackage{cite}
\usepackage{subcaption}
\usepackage{subfig} 
\usepackage{listings,lstlinebgrd}
\usepackage[data,amsfonts]{logic9-thm}
\usepackage{mathtools}
\usepackage{multirow}
\usepackage[colorinlistoftodos]{todonotes}
\usepackage{tikz}
\usepackage{expl3,xparse}

\usetikzlibrary{arrows,automata,backgrounds,calc,shapes,positioning}
\pgfdeclarelayer{background}
\pgfsetlayers{background,main}

\newcommand{\Rpos}{\ensuremath{\RR_{\ge 0}}}
\newcommand{\CConstr}[1]{\ensuremath{\Phi(#1)}}
\newcommand{\valinit}{\ensuremath{\mathbf{0}}}
\newcommand{\xra}[1]{\ensuremath{\xrightarrow{#1}}}
\newcommand{\tuple}[1]{\ensuremath{\langle #1 \rangle}}
\newcommand{\zg}{\ensuremath{\mathsf{ZG}}}
\newcommand{\abs}{\ensuremath{\mathfrak{a}}}
\newcommand{\extraLU}{\ensuremath{{\mathsf{Extra_{LU}}^+}}}
\newcommand{\simulated}{\ensuremath{\subseteq}}
\newcommand{\rank}{\ensuremath{\mathsf{rank}}}
\newcommand{\searchpo}{\ensuremath{\sqsubseteq}}

\newcommand{\topopo}{\ensuremath{\searchpo_{topo}}}

\DeclareCaptionFormat{listing}{%
  \rule{\dimexpr\textwidth\relax}{1pt}\par\vskip1pt#1#2#3
}
\lstdefinelanguage{algo}{%
  morekeywords={function,push,pop,top,for,all,or,not,if,then,else,repeat,until,while,do,report,return,such,that,pick,out,of,stack,list,int,order,descending,size,add,insert,remove,push_front,take,exit}
}

\ExplSyntaxOn
\NewDocumentCommand \lstcolorlines { O{green} m }
{
 \clist_if_in:nVT { #2 } { \the\value{lstnumber} }{ \color{#1} }
}
\ExplSyntaxOff

\begin{document}
\title{Improving search order for reachability testing in timed
  automata}
\author{Fr\'ed\'eric Herbreteau and Thanh-Tung Tran}
\institute{
Universit\'{e} de Bordeaux, Bordeaux INP, CNRS, LaBRI UMR5800\\
LaBRI B\^{a}t A30, 351 crs Lib\'{e}ration, 33405 Talence, France\\
}
\maketitle

\begin{abstract}
  Standard algorithms for reachability analysis of timed automata are
  sensitive to the order in which the transitions of the automata are
  taken. To tackle this problem, we propose a \emph{ranking system}
  and a \emph{waiting strategy}. This paper discusses the reason why
  the search order matters and shows how a ranking system and a
  waiting strategy can be integrated into the standard reachability
  algorithm to alleviate and prevent the problem
  respectively. Experiments show that the combination of the two
  approaches gives optimal search order on standard benchmarks except
  for one example. This suggests that it should be used instead of the
  standard BFS algorithm for reachability analysis of timed automata.
\end{abstract}

\section{Introduction}

Reachability analysis for timed automata asks if there is an execution
of an automaton reaching a given state.  This analysis can be used to
verify all kinds of safety properties of timed systems. The standard
approach to reachability analysis of timed automata uses sets of clock
valuations, called \emph{zones}, to reduce the reachability problem in
the infinite state space of a timed automaton to the reachability
problem in a finite graph. We present two heuristics to improve the
efficiency of the zone based reachability algorithm.

The algorithm for reachability analysis of timed automata is a
depth-first search, or a breadth-first search on a graph whose nodes
are pairs consisting of a state of the automaton and a zone describing
the set of possible clock valuations in this state. The use of zone
inclusion is crucial for efficiency of this algorithm. It permits to
stop exploration from a smaller zone if a bigger zone with the same
state has been already explored.

Due to the use of zone inclusion the algorithm is sometimes very
sensitive to exploration order. Indeed, it may happen that a small
zone is reached and explored first, but  then it is removed when a
bigger zone is reached later. We will refer to such a situation as a
\emph{mistake}. A mistake can often be avoided by taking a different
exploration order that reaches the bigger zone first.

In this paper we propose two heuristics to reduce the number of
mistakes in the reachability analysis. In the example below we explain
the mistake phenomenon in more details, and point out that it can
cause an exponential blowup in the search space; this happens in the
FDDI standard benchmark. The two heuristics are quite different in
nature, so we evaluate their performance on the standard
examples. Based on these experimental results we propose a simple
modification to the standard exploration algorithm that significantly
improves the exploration order.

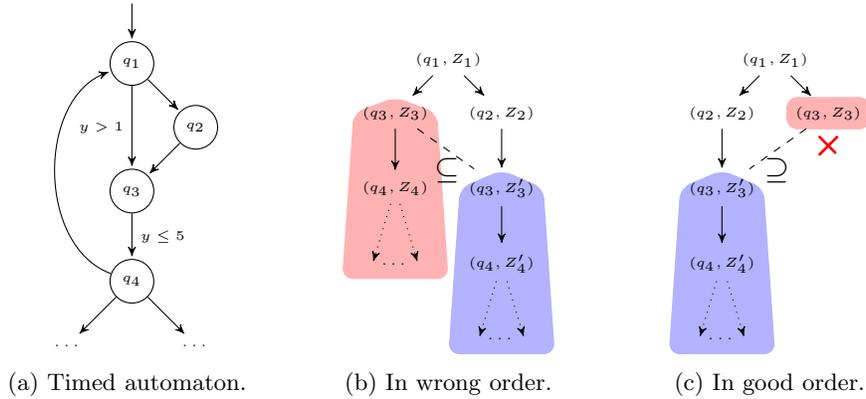
\begin{figure}[t]
  \begin{subfigure}[b]{.3\textwidth}
    \centering
    \begin{tikzpicture}[>=stealth',shorten >=1pt,auto,
      node distance=1.2cm, align=center]
      
      \tikzstyle{every node}=[font=\tiny]
      \tikzstyle{every state}=[minimum size = 10pt]
      
      \node[state]  (q1)                        {$q_1$};
      \node[state]  (q2)    [below right of=q1] {$q_2$};
      \node[state]  (q3)    [below left of=q2]  {$q_3$};
      \node[state]  (q4)    [below of=q3]       {$q_4$};
      \node         (qdot1) [below left of=q4]  {$\dots$};
      \node         (qdot2) [below right of=q4] {$\dots$};
      \path[->]
      ([yshift=0.5cm]q1.north) edge   (q1)
      (q1)  edge                       (q2)
      (q1)  edge node[left] {$y > 1$}  (q3)
      (q2)  edge                       (q3) 
      (q3)  edge node {$y \leq 5$}     (q4)
      (q4)  edge                       (qdot1) 
      (q4)  edge                       (qdot2)
      (q4)  edge[bend left=70]         (q1);
    \end{tikzpicture}
    \caption{Timed automaton.}
    \label{fig:racing}
  \end{subfigure}
  \hfill
  \begin{subfigure}[b]{0.3\textwidth}
    \centering
    \begin{tikzpicture}[>=stealth',shorten >=1pt,auto, node
      distance=1cm, align=center]
      \tikzstyle{every node}=[font=\tiny]
      \node (q1Z1)                           {$(q_1,Z_1)$};
      \node (q3Z3)    [below left of=q1Z1]   {$(q_3,Z_3)$};
      \node (q2Z2)    [below right of=q1Z1]  {$(q_2,Z_2)$};
      \node (q3Z3')   [below of=q2Z2]        {$(q_3,Z_3')$};
      \node (q4Z4)    [below of=q3Z3]        {$(q_4,Z_4)$};
      \node (q4Z4')   [below of=q3Z3']       {$(q_4,Z_4')$};
      \node (dots4)   [below of=q4Z4]        {$\dots$};
      \node (dots4')  [below of=q4Z4']       {$\dots$};

      \path[->]
      (q1Z1)  edge (q3Z3)
      (q1Z1)  edge (q2Z2)
      (q2Z2)  edge (q3Z3')
      (q3Z3)  edge (q4Z4)
      (q3Z3') edge (q4Z4');

      \path[->,dotted]
      (q4Z4)  edge (dots4.west)
      (q4Z4)  edge (dots4.east)
      (q4Z4') edge (dots4'.west)
      (q4Z4') edge (dots4'.east);

      \path[dashed]
      (q3Z3) edge
             node[below,font=\large] {$\simulated$}
      (q3Z3');

      \begin{pgfonlayer}{background}
        \filldraw[rounded corners,red!30]
        ([xshift=-.4cm,yshift=-.2cm]dots4.west) --
        (q3Z3.west) --
        (q3Z3.north) --
        (q3Z3.east) --
        ([xshift=.4cm,yshift=-.2cm]dots4.east) --
        cycle;
        \filldraw[rounded corners,blue!30]
        ([xshift=-.4cm,yshift=-.2cm]dots4'.west) --
        (q3Z3'.west) --
        (q3Z3'.north) --
        (q3Z3'.east) --
        ([xshift=.4cm,yshift=-.2cm]dots4'.east) --
        cycle;
      \end{pgfonlayer}
    \end{tikzpicture}
    \caption{In wrong order.}
    \label{fig:exploration}
  \end{subfigure}
  \hfill
  \begin{subfigure}[b]{0.3\textwidth}
    \centering
    \begin{tikzpicture}[>=stealth',shorten >=1pt,auto, node
      distance=1cm, align=center]
      \tikzstyle{every node}=[font=\tiny]
      \node (q1Z1)                           {$(q_1,Z_1)$};
      \node (q2Z2)    [below left of=q1Z1]   {$(q_2,Z_2)$};
      \node (q3Z3)    [below right of=q1Z1]  {$(q_3,Z_3)$};
      \node (q3Z3')   [below of=q2Z2]        {$(q_3,Z_3')$};
      \node (q4Z4')   [below of=q3Z3']       {$(q_4,Z_4')$};
      \node (dots4')  [below of=q4Z4']       {$\dots$};

      \path[->]
      (q1Z1)  edge (q2Z2)
      (q1Z1)  edge (q3Z3)
      (q2Z2)  edge (q3Z3')
      (q3Z3') edge (q4Z4');

      \path[->,dotted]
      (q4Z4') edge (dots4'.west)
      (q4Z4') edge (dots4'.east);

      \node[draw=red,cross out,line width=1pt]
      (cross) [below=0.1cm of q3Z3] {};

      \path[dashed]
      (q3Z3) edge
             node[below,font=\large] {$\supseteq$}
      (q3Z3');

      \begin{pgfonlayer}{background}
        \filldraw[rounded corners,red!30]
        (q3Z3.north west) --
        (q3Z3.north east) --
        (q3Z3.south east) --
        (q3Z3.south west) --
        cycle;
        \filldraw[rounded corners,blue!30]
        ([xshift=-.4cm,yshift=-.2cm]dots4'.west) --
        (q3Z3'.west) --
        (q3Z3'.north) --
        (q3Z3'.east) --
        ([xshift=.4cm,yshift=-.2cm]dots4'.east) --
        cycle;
      \end{pgfonlayer}
    \end{tikzpicture}
    \caption{In good order.}
    \label{fig:subsumption}
  \end{subfigure}
  \caption{A timed automaton and two exploration graphs of its
    state-space. On the left, the transition to $q_3$ is explored
    first, which results in exploring the subtree of $q_3$ twice. On
    the right, the transition to $q_2$ is explored first and
    subsumption stops the second exploration as $Z_3$ is included in
    $Z'_3$.}
  \label{fig:racing-situation}
\end{figure}


We now give a concrete example showing why exploration order
matters. Consider the timed automaton shown in Figure~\ref{fig:racing},
and assume we perform a depth-first search (DFS) exploration of its
state space. The algorithm starts in $(q_1,Z_1)$ where
$Z_1 = (y \ge 0)$ is the set of all clock values. Assume that the
transition to $q_3$ is taken first as in
Figure~\ref{fig:exploration}. The algorithm reaches the node
$(q_3, Z_3)$ with $Z_3 = (y > 1)$ and explores its entire
subtree. Then, the algorithm backtracks to $(q_1, Z_1)$ and proceeds
with the transition to $q_2$ reaching $(q_2, Z_2)$, and then
$(q_3, Z_3')$ with $Z_2 = Z_3' = (y \ge 0)$. It happens that
$Z_3 \subseteq Z_3'$: the node $(q_3, Z_3')$ is \emph{bigger} than the
node $(q_3, Z_3)$ which has been previously visited. At this point,
the algorithm has to visit the entire subtree of $(q_3, Z_3')$ since
the clock valuations in $Z_3' \setminus Z_3$ have not been
explored. The net result is that the earlier exploration from
$(q_3, Z_3)$ turns out to be useless since we need to explore from
$(q_3, Z_3')$ anyway. If, by chance, our DFS exploration had taken
different order of transitions, and first considered the one from
$q_1$ to $q_2$ as in Figure~\ref{fig:subsumption}, the exploration would
stop at $(q_3, Z_3)$ since the bigger node $(q_3, Z_3')$ has already
been visited and $Z_3 \subseteq Z_3'$. To sum up, in some cases DFS
exploration is very sensible to the search order.

Several authors~\cite{BHV00, BEH05} have observed that BFS exploration
is often much more efficient than DFS for reachability testing in
timed automata. This can be attributed to an empirical observation
that often a zone obtained by a short path is bigger than the one
obtained by a longer path. This is the opposite in our example from
Figure~\ref{fig:racing}. In consequence, a BFS algorithm will also do
unnecessary explorations. When $(q_3, Z_3')$ is visited, the node
$(q_4, Z_4)$ is already in the queue. Hence, while the algorithm has a
chance to realise that exploring $(q_3, Z_3)$ is useless due to the
bigger node $(q_3, Z_3')$, it will keep visiting $(q_4, Z_4)$ and all
the subtree of $(q_3, Z_3)$. Indeed, in the standard BFS algorithm,
there is no mechanism to remove $(q_4, Z_4)$ from the queue when
$(q_3, Z_3')$ is reached. Again, considering the transition from $q_1$
to $q_2$ before the transition to $q_3$ as in
Figure~\ref{fig:subsumption}, avoids unnecessary exploration. Yet, by
making the path $q_1 \to q_2 \to q_3$ one step longer we would obtain
an example where all choices of search order would lead to unnecessary
exploration. Overall, the standard reachability algorithm for timed
automata, be it DFS or BFS, is sensitive to the alignment between the
discovery of big nodes and the exploration of small nodes.

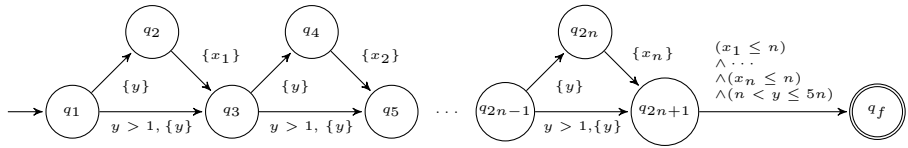
\begin{figure}[h]
  \centering
  \begin{tikzpicture}[>=stealth',shorten >=1pt,auto, node
    distance=1.5cm, align=center]
    \tikzstyle{every node}=[font=\tiny]
    \tikzstyle{every state}=[minimum size = 20pt]
    \node[state]  (q1)                           {$q_1$};
    \node[state]  (q2)    [above right of=q1]    {$q_2$};
    \node[state]  (q3)    [below right of=q2]    {$q_3$};
    \node[state]  (q4)    [above right of=q3]    {$q_4$};
    \node[state]  (q5)    [below right of=q4]    {$q_5$};
    \node         (qdot)  [right=0.1cm of q5]    {$\ldots$};
    \node[state,inner sep=0.5pt]
                  (q2nm1) [right=0.1cm of qdot]  {$q_{2n-1}$};
    \node[state]  (q2n)   [above right of=q2nm1] {$q_{2n}$};
    \node[state,inner sep=2pt]
                  (q2np1) [below right of=q2n]   {$q_{2n+1}$};
    \node[state,accepting,node distance=1.6cm]
                  (qf)    [right=2cm of q2np1]   {$q_f$};
    
    \path[->]
    ([xshift=-0.5cm]q1.west) edge                          (q1)
    (q1)    edge node[swap]  {$\{y\}$}                     (q2)
    (q1)    edge node[below] {$y > 1,\{y\}$}               (q3) 
    (q2)    edge node        {$\{x_1\}$}                   (q3)
    (q3)    edge node[swap]  {$\{y\}$}                     (q4)
    (q3)    edge node[below] {$y > 1$, $\{y\}$}            (q5)
    (q4)    edge node        {$\{x_2\}$}                   (q5)
    (q2nm1) edge node[swap]  {$\{y\}$}                     (q2n)
    (q2nm1) edge node[below] {$y > 1$,$\{y\}$}             (q2np1) 
    (q2n)   edge node        {$\{x_n\}$}                   (q2np1)
    (q2np1) edge node[above] {$\begin{array}{l}
                                 (x_1 \le n) \\
                                 \land \cdots \\
                                 \land (x_n \le n) \\
                                 \land (n < y \le 5n)
                               \end{array}$} (qf);
  \end{tikzpicture}
  \caption{Timed automaton with a racing situation.}
  \label{fig:blowup}
\end{figure}


One could ask what can be the impact of a pattern from
Figure~\ref{fig:racing}, and does it really occur. The blowup of the
exploration space can be exponential. One example is presented in
Figure~\ref{fig:blowup}. It is obtained by iterating $n$ times the
pattern we have discussed above. The final state $q_f$ is not
reachable. By a similar analysis we can show that both the BFS and DFS
algorithms with wrong exploration order explore and store
exponentially more nodes than needed. In the automaton there are $2^n$
different paths to $q_{2n+1}$. The longest path
$q_1,q_2,q_3,\dots,q_{2n+1}$ generates the biggest zone, while there
are about $2^n$ different zones that can be generated by taking
different paths. If the DFS takes the worst exploration order, all
these zones will be generated. If it takes the wrong order half of the
times, then about $2^{n/2}$ zones will be generated. Similarly for
BFS.

In the experiments section we show that, this far from optimal
behaviour of BFS and DFS exploration indeed happens in the FDDI model,
a standard benchmark model for timed automata.

In this paper we propose simple modifications of the exploration
strategy to counter the problem as presented in the above examples. We
will first describe a \emph{ranking system} that mitigates the problem
by assigning ranks to states, and using ranks to chose the transitions
to explore. It will be rather clear that this system addresses the
problem from our examples. Then we will propose \emph{waiting
  strategy} that starts from a different point of view and is simpler
to implement. The experiments on standard benchmarks show that the two
approaches are incomparable but they can be combined to give optimal
results in most of the cases. Since this combination is easy to
implement, we propose to use it instead of standard BFS for
reachability checking.

\paragraph{Related work:} The influence of the search order has been
discussed in the literature in the context of state-caching
\cite{DBLP:journals/fmsd/GodefroidHP95,DBLP:conf/cav/BehrmannLP03,pelanek:report:2008,DBLP:journals/topnoc/EvangelistaK10},
and state-space fragmentation\cite{BHV00,BEH05,BOS02}. State-caching
focuses on limiting the number of stored nodes at the cost of
exploring more nodes. We propose a strategy that improves the number
of visited nodes as well as the number of stored
nodes. In\cite{BHV00,BEH05,BOS02}, it is suggested that BFS is the
best search order to avoid state-space fragmentation in distributed
model checking. We have not yet experimented our approach for
distributed state-space exploration.

In terms of implementation, our approaches add a metric to states. In
a different context a metric mechanism has been used by Behrmann
\emph{et al.} to guide the exploration in priced timed automata in
\cite{BF01}.

\paragraph{Organisation of the paper:}
In the next section we present preliminaries for this paper: timed
automata, the reachability problem and the standard reachability
algorithm for timed automata. In Section~\ref{sec:ranking}, we propose
a ranking system to limit the impact of mistakes during
exploration. Section~\ref{sec:joint} presents another strategy that
aims at limiting the number of mistakes. Finally,
Section~\ref{sec:experiments} gives some experimental results on the
standard benchmarks.


\section{Preliminaries}
\label{sec:preliminaries}

We introduce preliminary notions about timed automata and the
reachability problem. Then, we introduce the classical zone-based
algorithm used to solve this problem.

\subsection{Timed Automata and the Reachability Problem}
\label{sec:preliminaries:ta-reach}

Let $X = \{ x_1, \dots, x_n \}$ be a set of clocks, i.e. variables
that range over the non-negative real numbers $\Rpos$.
A \emph{clock constraint} $\phi$ is a conjunction of constraints
$x \# c$ for $x \in X$, $\# \in \{ <, \le, =, \ge, > \}$ and
$c \in \Nat$. Let $\CConstr{X}$ be the set of clock constraints over
the set of clocks $X$.
A \emph{valuation} over $X$ is a function $v: X \to \Rpos$. We denote
by $\valinit$ the valuation that maps each clock in $X$ to $0$, and by
$\Rpos^X$ the set of valuations over $X$.
A valuation $v$ satisfies a clock constraint $\phi \in \CConstr{X}$,
denoted $v \models \phi$, when all the constraints in $\phi$ hold
after replacing every clock $x$ by its value $v(x)$.
For $\delta \in \Rpos$, we denote $v + \delta $ the valuation that
maps every clock $x$ to $v(x) + \delta$. For $R \subseteq X$, $R[v]$
is the valuation that sets $x$ to $0$ if $x \in R$, and that sets $x$
to $v(x)$ otherwise.

A \emph{timed automaton (TA)} is a tuple
$\Aa = (Q, q_0, F, X, Act, T)$ where $Q$ is a finite set of states
with initial state $q_0 \in Q$ and accepting states $F \subseteq Q$,
$X$ is a finite set of clocks, $Act$ is a finite alphabet of actions,
$T \subseteq Q \times \CConstr{X} \times 2^X \times Act \times Q$ is a
finite set of transitions $(q, g, R, a, q')$ where $g$ is a
\emph{guard}, $R$ is the set of clocks that are \emph{reset} and $a$
is the \emph{action} of the transition.

The semantics of a TA $\Aa$ is given by a transition system whose
states are \emph{configurations} $(q,v) \in Q \times \Rpos^X$. The
\emph{initial configuration} is $(q_0,\valinit)$. We have delay
transitions: $(q,v) \xra{\d} (q, v+\d)$ for $\d \in \Rpos$, and action
transitions: $(q,v) \xra{a} (q',v')$ if there exists a transition
$(q, g, R, a, q') \in T$ such that $v \models g$ and $v' = [R] v$.
A \emph{run} is a finite sequence of transitions starting from the
initial configuration $(q_0,\valinit)$. A run is \emph{accepting} is
it ends in a configuration $(q,v)$ with an accepting state $q \in F$.

The \emph{reachability problem} consists in deciding if a given TA
$\Aa$ has an accepting run. This problem is known to be
$\PSPACE$-complete\cite{AD94}.

\subsection{Symbolic Semantics}
\label{sec:preliminaries:symbolic}

The reachability problem cannot be solved directly from $\Aa$ due to
the uncountable number of configurations. The standard solution is to
use symbolic semantics of timed automata by grouping valuations
together. A \emph{zone} is a set of valuations described by a
conjunction of two kinds of constraints: $x_i \# c$ and
$x_i - x_j \# c$ where $x_i,x_j \in X$, $c\in \ZZ$ and
$\# \in \set{<, \le, =, \ge, >}$.

The \emph{zone graph} $\zg(\Aa)$ of a timed automaton
$\Aa = (Q, q_0, F, X, Act, T)$ is a transition system with nodes of
the form $(q, Z)$ where $q \in Q$ and $Z$ is a zone. The initial node
is $(q_0, Z_0)$ where
$Z_0 = \set{ \valinit + \delta \mid \delta \in \Rpos}$. The nodes
$(q,Z)$ with $q \in F$ are accepting. There is a transition
$(q, Z) \Rightarrow (q', Z')$ if there exists a transition
$(q, g, R, a, q') \in T$ such that
$Z' = \set{ v' \in \Rpos \mid \exists v \in Z \, \exists \d \in \Rpos
  \ (q,v) \xra{a} \xra{\d} (q',v') }$ and $Z' \neq \es$.
The relation $\Rightarrow$ is well-defined as it can be shown that if
$Z$ is a zone, then $Z'$ is a zone. Zones can be efficiently
represented by Difference Bound Matrices (DBMs)~\cite{Dil89} and the
successor $Z'$ of a zone $Z$ can be efficiently computed using this
representation.

The zone graph $\zg(\Aa)$ is still infinite~\cite{DT98}, and an
additional abstraction step is needed to obtain a finite transition
system. An \emph{abstraction operator} is a function
$\abs : \Pp(\Rpos^X) \to \Pp(\Rpos^X)$ such that $W \subseteq \abs(W)$
and $\abs(\abs(W)) = \abs(W)$ for every set $W$ of valuations. An
abstraction operator defines an abstract symbolic semantics.
Similarly to the zone graph, we define the \emph{abstract zone graph}
$\zg^\abs(\Aa)$. Its initial node is $(q_0, \abs(Z_0))$ and we have a
transition $(q,Z) \Rightarrow_{\abs} (q',\abs(Z'))$ if $\abs(Z) = Z$
and $(q,Z) \Rightarrow (q',Z')$.

In order to solve the reachability problem for $\Aa$ from
$\zg^\abs(\Aa)$, the abstraction operator $\abs$ should have the
property that every run of $\Aa$ has a corresponding path in
$\zg^\abs(\Aa)$ (completeness) and conversely, every path in
$\zg^\abs(\Aa)$ should correspond to a run in $\Aa$
(soundness). Furthermore, $\zg^\abs(\Aa)$ should be finite. Several
abstraction operators have been introduced in the
literature\cite{DT98,BBLP06}. The abstraction operator
$\extraLU$\cite{BBLP06} has all the required properties
above. Moreover, the $\extraLU$ abstraction of a zone is itself a
zone. It can be computed from the DBM representation of the zone. This
allows to compute the abstract zone graph efficiently using DBMs as a
symbolic representation for zones. The $\extraLU$ abstraction is used
by most implementation including the state-of-the-art tool
UPPAAL\cite{Behrmann:QEST:2006}. The theorem below reduces the
reachability problem for $\Aa$ to the reachability problem in the
finite graph $\zg^{\extraLU}(\Aa)$.

\begin{theorem}[\cite{BBLP06}]
  There is an accepting run in $\Aa$ iff there exists a path in
  $\zg^\extraLU(\Aa)$ from $(q_0, \extraLU(Z_0))$ to some state $(q,Z)$
  with $q \in F$. Furthermore $\zg^\extraLU(\Aa)$ is finite.
\end{theorem}

\subsection{Reachability algorithm}
\label{sec:preliminaries:algorithm}

\begin{lstlisting}[frame=bt, caption={Standard reachability algorithm
    for timed automaton $\Aa$.}, label={algo:standard-reachability},
  float]
function €\textsf{reachability\_check}€($\Aa$)
  $W$ := $\set{ (q_0, \extraLU(Z_0)) }$; $P$ := $W$ €// Invariant: $W
  \subseteq P$€

  while $(W \neq \es)$ do
    take and remove a node $(q, Z)$ from $W$€\label{algo:standard-reachability:search-order}€
    if ($q$ is accepting)
      return Yes
    else 
      for each $(q, Z) \Rightarrow_\extraLU (q', Z')$
        if there is no $(q_B, Z_B) \in P$ s.t. $(q', Z') \simulated (q_B, Z_B)$€\label{algo:standard-reachability:simulated}€
          for each  $(q_S, Z_S)\in P$ such that $(q_S, Z_S) \simulated (q', Z')$ 
             remove  $(q_S, Z_S)$ from $W$ and $P$€\label{algo:standard-reachability:maximal}€
          add $(q', Z')$ to $W$ and to $P$

  return No
\end{lstlisting}

Algorithm~\ref{algo:standard-reachability} is the standard
reachability algorithm for timed automata. It explores the finite
abstract zone graph $\zg^{\extraLU}(\Aa)$ of an automaton $\Aa$ from
the initial node until it finds an accepting node, or it has visited
the entire state-space of $\zg^{\extraLU}(\Aa)$. It maintains a set of
\emph{waiting nodes} $W$ and a set of \emph{visited nodes} $P$ such
that $W \subseteq P$.

Algorithm~\ref{algo:standard-reachability} uses zone inclusion to stop
exploration, and this is essential for its efficiency. We have
$(q, Z) \subseteq (q', Z')$ when $q=q'$ and $Z \subseteq Z'$.  Notice
that zone inclusion is a simulation relation over nodes since zones are
sets of valuations. Zone inclusion is first used in
line~\ref{algo:standard-reachability:simulated} to stop the
exploration in $(q,Z)$ if there is a bigger node $(q_B, Z_B)$ in
$P$. It is also used in line~\ref{algo:standard-reachability:maximal}
to only keep the maximal nodes w.r.t. $\simulated$ in $P$ and $W$.

Algorithm~\ref{algo:standard-reachability} does not specify any
exploration strategy. As we have stressed in the introduction, the search order greatly
influences the number of nodes visited by the algorithm and stored in
the sets $W$ and $P$. At first sight it may seem strange why there
should be a big difference between, say, BFS and DFS search
orders. The cause is the optimisation due to subsumption
w.r.t. $\simulated$ in
lines~\ref{algo:standard-reachability:simulated}
and~\ref{algo:standard-reachability:maximal}. When equality on nodes
is used instead of zone inclusion, every node is visited. Hence, BFS
and DFS coincide in the sense that they will visit the same nodes,
while not in the same order. The situation is very different with zone
inclusion. Consider again the two nodes
$(q_2, Z_2) \simulated (q_2, Z_2')$ in
Figure~\ref{fig:exploration}. Since the smaller node $(q_2, Z_2)$ is
reached first, the entire subtrees of both nodes are visited whereas
it would be sufficient to explore the subtree of the bigger node
$(q_2,Z_2')$ to solve the reachability problem. Indeed, every node
below $(q_2,Z_2)$ is simulated by the corresponding node below
$(q_2,Z_2')$. Notice that the problem occurs both with a DFS and with
a BFS strategy since the bigger node $(q_2,Z_2')$ is further from the
root node than the smaller node $(q_2, Z_2)$. When the bigger node is
found before the smaller one, as in Figure~\ref{fig:subsumption}, only
the subtree of the bigger node is visited. An optimal search strategy
would guarantee that big nodes are visited before small ones. In the
remaining of the paper we propose two heuristics to optimise the
search order.


\section{Ranking system}
\label{sec:ranking}

In this section we propose an exploration strategy to address the
phenomenon we have presented in the introduction: we propose a
solution to stop the exploration of the subtree of a small node when a
bigger node is reached. As we have seen, the late discovery of big
nodes causes unnecessary explorations of small nodes and their
subtrees. In the worst case, the number of needlessly visited nodes
may be exponential (cf. Figure~\ref{fig:blowup}).

Our goal is to minimise the number of visited nodes as well as the
number of stored nodes (i.e. the size of $P$ in
Algorithm~\ref{algo:standard-reachability}). Consider again the
situation in Figure~\ref{fig:exploration} where
$(q_3, Z_3) \simulated (q_3, Z_3')$. When the big node $(q_3, Z_3')$
is reached, we learn that exploring the small node $(q_3, Z_3)$ is
unnecessary. In such a situation,
Algorithm~\ref{algo:standard-reachability} erases the small node
$(q_3, Z_3)$ (line~\ref{algo:standard-reachability:simulated}), but
all its descendants that are in the waiting list $W$ will be still
explored.

A first and straightforward solution would be to erase the whole
subtree of the small node $(q_3, Z_3)$.
Algorithm~\ref{algo:standard-reachability} would then proceed with the
waiting nodes in the subtree of $(q_3, Z_3')$.  This approach is
however too rudimentary. Indeed, it may happen that the two nodes
$(q_4, Z_4)$ and $(q_4, Z_4')$ in Figure~\ref{fig:exploration} are
identical. Then, erasing the whole subtree of $(q_3, Z_3)$ will lead
to exploring $(q_4, Z_4)$ and all its subtree twice. We have observed
on classical benchmarks (see Section~\ref{sec:experiments}) that
identical nodes are frequently found. While this approach is correct,
it would result in visiting more nodes than the classical algorithm.

We propose a more subtle approach based on an interesting property of
Algorithm~\ref{algo:standard-reachability}. Consider the two nodes
$(q_4, Z_4)$ and $(q_4, Z_4')$ in Figure~\ref{fig:exploration} again,
and assume that $(q_4, Z_4')$ is reached after $(q_4, Z_4)$. If the
two nodes are identical, then $(q_4, Z_4')$ is erased by
Algorithm~\ref{algo:standard-reachability} in
line~\ref{algo:standard-reachability:simulated}, but $(q_4, Z_4)$ is
kept since it has been visited first. Conversely, if the two nodes are
different, we still have $(q_4, Z_4) \simulated (q_4, Z_4')$, then
$(q_4, Z_4)$ is erased by Algorithm~\ref{algo:standard-reachability}
in line~\ref{algo:standard-reachability:simulated}. Hence, as the
algorithm explores the subtree of $(q_3, Z_3')$, it progressively
erases all the nodes in the subtree of $(q_3, Z_3)$ that are smaller
than some node in the subtree of $(q_3, Z_3')$. At the same time, it
keeps the nodes that are identical to some node below $(q_3, Z_3')$,
hence avoiding several explorations of the same node.

Now, it remains to make all this happen before the subtree of
$(q_3, Z_3)$ is developed any further. This is achieved by giving a
higher priority to $(q_3, Z_3')$ than all the waiting nodes below
$(q_3, Z_3)$. This priority mechanism is implemented by assigning a
\emph{rank} to every node. 

Algorithm~\ref{algo:ranking-reachability}
below is a modified version of
Algorithm~\ref{algo:standard-reachability} that implements the ranking
of nodes (the modifications are highlighted). Nodes are initialised
with rank $0$. The rank of a node $(q', Z')$ is updated with respect
to the ranks of the nodes $(q_S, Z_S)$ that are simulated by
$(q', Z')$ (line~\ref{algo:ranking-reachability:update-rank}). For
each node $(q_S, Z_S)$, we compute the maximum rank $r$ of the waiting
nodes below $(q_S, Z_S)$. Then, $\rank(q',Z')$ is set to
$\max(\rank(q',Z'), r+1)$ giving priority to $(q',Z')$ over the
waiting nodes below $(q_S, Z_S)$.

\begin{lstlisting}[frame=bt, caption={Reachability algorithm with
    ranking of nodes for timed automaton $\Aa$. The set $P$ is stored
    as a tree $\rightarrow$.},
  label={algo:ranking-reachability},basicstyle=\scriptsize,
  linebackgroundcolor={\lstcolorlines[orange!30]{3,6,11,14,15,20,21,22,23,24,25,26,27,28,29,30,31,32,33}}]
function €\textsf{reachability\_check}€($\Aa$)
  $W$ := $\set{ (q_0, \extraLU(Z_0)) }$; $P$ := $W$
  $\mathsf{init\_rank}(q_0, \extraLU(Z_0))$

  while $(W \neq \es)$ do
    take and remove a node $(q, Z)$ €\textbf{with highest rank}€ from $W$€\label{algo:ranking-reachability:search-order}€
    if ($q$ is accepting) then
      return Yes
    else 
      for each $(q, Z) \Rightarrow_\extraLU (q', Z')$
        $\mathsf{init\_rank}(q',Z')$
        if there is no $(q_B, Z_B) \in P$ s.t. $(q', Z') \simulated (q_B, Z_B)$ then €\label{algo:ranking-reachability:simulated}€
          for each $(q_S, Z_S) \in P$ s.t. $(q_S, Z_S) \simulated (q',Z')$
            if $(q_S,Z_S) \not \in W$ then // implies not a leaf node in $P$
              $\rank(q', Z')$ := $\max(\rank(q',Z'), \ 1+\mathsf{max\_rank\_waiting}(q_S,Z_S))$€\label{algo:ranking-reachability:update-rank}€
            remove  $(q_S, Z_S)$ from $W$ and $P$€\label{algo:ranking-reachability:maximal}€
          add $(q', Z')$ to $W$ and to $P$€\label{algo:ranking-reachability:adding-to-W-P}€
  return No

function €\textsf{max\_rank\_waiting}€($q,Z$)
  if $(q, Z)$ is in $W$ then // implies leaf node in $P$
    return $\rank(q,Z)$
  else
    $r$ := 0;
    for each edge $(q,Z) \rightarrow (q',Z')$ in $P$
      $r$ := $\max(r, \text{max\_rank\_waiting}(q',Z'))$
    return $r$

function €\textsf{init\_rank}€($q,Z$)
  if $Z$ is the $true$ zone then
    $\rank(q,Z)$ := $\infty$€\label{algo:ranking-reachability:rank-true-zone}€
  else
    $\rank(q,Z)$ := $0$
\end{lstlisting}

The function \textsf{max\_rank\_waiting} determines the maximal rank
among waiting nodes below $(q_S,Z_S)$. To that purpose, the set of
visited nodes $P$ is stored as a reachability tree. When a node
$(q_S,Z_S)$ is removed in
line~\ref{algo:ranking-reachability:maximal}, its parent node is
connected to its child nodes to maintain reachability of waiting
nodes. Observe that the node $(q',Z')$ is added to the tree $P$ in
line~\ref{algo:ranking-reachability:adding-to-W-P} after its rank has
been updated in line~\ref{algo:ranking-reachability:update-rank}. This
is needed in the particular case where $(q_S, Z_S)$ is an ancestor of
node $(q',Z')$ in
line~\ref{algo:ranking-reachability:update-rank}. The rank of
$(q',Z')$ will be updated taking into account the waiting nodes below
$(q_S,Z_S)$. Obviously, $(q',Z')$ should not be considered among those
waiting nodes, which is guaranteed since $(q',Z')$ does not belong to
the tree yet.

The intuition behind the use of ranks suggest one more useful
heuristic. Ranks are used to give priority to exploration from some
nodes over the others. Nodes with true zones are a special case in
this context, since they can never be covered, and in consequence it
is always better to explore them first. We implement this observation
by simply assigning the biggest possible rank ($\infty$) to such nodes
(line~\ref{algo:ranking-reachability:rank-true-zone} in the
Algorithm).

Let us explain how the Algorithm~\ref{algo:ranking-reachability} works
on an example. Consider again the automaton in
Figure~\ref{fig:racing}.  The final exploration graph is depicted in
Figure~\ref{fig:stopRacing}. When $(q_1,Z_1)$ is visited, both
$(q_3,Z_3)$ and $(q_2,Z_2)$ are put into the waiting list $W$ with
rank $0$. Recall that exploring $(q_3, Z_3)$ first is the worst
exploration order. This adds $(q_4,Z_4)$ to the waiting list with rank
$0$. The exploration of $(q_2,Z_2)$ adds $(q_3,Z_3')$ to the waiting
list. At this stage, the rank of $(q_3,Z_3')$ is set to $1$ since it
is bigger than $(q_3,Z_3)$ which is erased. The node $(q_3,Z_3')$ has
the highest priority among all waiting nodes and is explored
next. This generates the node $(q_4,Z_4')$ that is bigger than
$(q_4,Z_4)$. Hence $(q_4, Z_4)$ is erased, $(q_4, Z_4')$ gets rank $1$
and the exploration proceeds from $(q_4,Z_4')$. One can see that, when
a big node is reached, the algorithm not only stops the exploration of
the smaller node but also of the nodes in its
subtree. Figure~\ref{fig:stopRacing} shows a clear improvement over
Figure~\ref{fig:exploration}.

\begin{figure}[t]
  \centering
  \begin{minipage}[b]{0.4\textwidth}
    \centering
    \begin{tikzpicture}[>=stealth',shorten >=1pt,auto, node
      distance=.4cm, align=center]
      \tikzstyle{every node}=[font=\tiny]
      \node (q1Z1)                          {$(q_1,Z_1)$};
      \node (q3Z3)   [below left=of q1Z1]   {$(q_3,Z_3)$};
      \node (q2Z2)   [below right=of q1Z1]  {$(q_2,Z_2)$};
      \node (q4Z4)   [below=of q3Z3]        {$(q_4,Z_4)$};
      \node (q3Z3')  [below=of q2Z2]        {$(q_3,Z_3')$};
      \node (q4Z4')  [below=of q3Z3']       {$(q_4,Z_4')$};
      \node (dots)   [below=of q4Z4']       {$\dots$};
      
      \path[->]
      (q1Z1)  edge (q3Z3)
      (q1Z1)  edge (q2Z2)
      (q3Z3)  edge (q4Z4)
      (q2Z2)  edge (q3Z3')
      (q3Z3') edge (q4Z4');
      
      \path[->,dotted]
      (q4Z4') edge (dots.west)
      (q4Z4') edge (dots.east);
      
      \node[draw=red,cross out,line width=1pt]
      (cross) [below=0.1cm of q4Z4] {};
      
      \path[dashed,line width=1pt]
      (q3Z3) edge node[below,font=\large] {$\simulated$} (q3Z3')
      (q4Z4) edge node[below,font=\large] {$\simulated$} (q4Z4');
      
      \begin{pgfonlayer}{background}
        \filldraw[rounded corners,blue!30]
        ([xshift=-.4cm]dots.south west) --
        (q3Z3'.north west) --
        (q3Z3'.north) --
        (q3Z3'.north east) --
        ([xshift=.4cm]dots.south east) --
        cycle;
        
        \filldraw[rounded corners,red!30]
        (q4Z4.south west) --
        (q3Z3.north west) --
        (q3Z3.north east) --
        (q4Z4.south east) --
        cycle;
      \end{pgfonlayer}
    \end{tikzpicture}
    \caption{Reachability tree for
      Algorithm~\ref{algo:ranking-reachability} on the automaton in
      Figure~\ref{fig:racing}.}
    \label{fig:stopRacing}
  \end{minipage}
  \hfill
  \begin{minipage}[b]{0.45\textwidth}
    \centering
    \begin{tikzpicture}[>=stealth',shorten >=1pt,auto, node
      distance=1cm, align=center]
      \tikzstyle{every node}=[font=\footnotesize]
      \node (root) {};
      
      \node (S) [below=1cm of root]          {$s$};
      \node (sdot1) [below left of=S]        {$\dots$};
      \node (sdot2) [right=.1cm of sdot1 ]   {$\dots$};
      \node (sdot3) [below right of=S]       {$\dots$};
      
      \node (V1) [below left of=sdot1]       {$v_1$};
      \node (V2) [right=.2cm of V1]          {$v_2$};
      \node (Vn) [below right of=sdot3]      {$v_n$};
      
      \node (tdot1) [below right of=V1]      {$\dots$};
      \node (tdot2) [right=.1cm of tdot1]    {$\dots$};
      \node (tdot3) [below left of=Vn]       {$\dots$};
      
      \node (t) [below right of=tdot1]       {$\mathbf{t}$};

      \node (dots1) [below left of=t]        {$\dots$};
      \node (dots2) [below right of=t]       {$\dots$};

      \path[->]
      (S)      edge (sdot1)
      (S)      edge (sdot2)
      (S)      edge (sdot3)
      (sdot1)  edge (V1)
      (sdot2)  edge (V2)
      (sdot3)  edge (Vn)
      (V1)     edge (tdot1)
      (V2)     edge (tdot2)
      (Vn)     edge (tdot3)
      (tdot1)  edge (t)
      (tdot2)  edge (t)
      (tdot3)  edge (t);

      \path[->,dotted]
      (t) edge (dots1.west)
      (t) edge (dots1.east)
      (t) edge (dots2.west)
      (t) edge (dots2.east);

      \path[dotted]
      ([xshift=.2cm] V2.east) edge ([xshift=-.2cm] Vn.west);
    \end{tikzpicture}
    \caption{Waiting strategy starts exploring from $t$ only after all
      paths leading to $t$ are explored.}
    \label{fig:diamond}
  \end{minipage}
\end{figure}



\section{Waiting strategy}
\label{sec:joint}

We present an exploration strategy that will aim at reducing the
number of exploration mistakes: situations when a bigger node is
discovered later than a smaller one.  The ranking strategy from the
previous section reduced the cost of a mistake by stopping
the exploration from descendants of a small node when it found a bigger
node. By contrast, the waiting strategy of this section
will not develop a node if it is aware of some other parts of
exploration that may lead to a bigger node.

The waiting strategy is based on topological-like order on states of
automata. We first present this strategy on a single automaton. Then
we consider networks of timed automata, and derive a topological-like
ordering from orderings on the components. Before we start we explain
what kind of phenomenon our strategy is capturing.

To see what we aim at, consider the part of a timed automaton depicted
in Figure~\ref{fig:diamond}. There is a number of paths form state $s$
to state $t$, not necessary of the same length. Suppose the search
strategy from $(s,Z)$ has reached $(t,Z_1)$ by following the path
through $v_1$. At this point it is reasonable to delay exploration
from $(t,Z_1)$ until all explorations of paths through $v_2,\dots,v_k$
finish. This is because some of these explorations may result in a
bigger zone than $Z_1$, and in consequence make an exploration from
$(t,Z_1)$ redundant.

The effect of such a waiting heuristic is clearly visible on our
example from Figure~\ref{fig:blowup}. The automaton consists of
segments: from $q_1$ to $q_3$, from $q_3$ to $q_5$, etc. Every segment
is a very simple instance of the situation from
Figure~\ref{fig:diamond} that we have discussed in the last
paragraph. There are two paths that lead from state $q_1$ to state
$q_3$. These two paths have different lengths, so with a BFS
exploration one of the paths will reach $q_3$ faster than the
other. The longest path (that one going through $q_2$) gives the
biggest zone in $q_3$; but BFS will no be able to use this
information; and in consequence it will generate exponentially many
nodes on this example.  The waiting heuristic will collect all the
search paths at states $q_3,q_5,\dots$ and will explore only the best
ones, so its search space will be linear.

We propose to implement these ideas via a simple modification of the
standard algorithm. The waiting strategy will be based on a partial
order $\topopo$ of sates of $\Aa$. We will think of it as a
topological order of the graph of the automaton (after removing cycles
in some way). This order is then used to determine the exploration
order.

\begin{lstlisting}[frame=bt, caption={Reachability algorithm with
    waiting strategy}, label={algo:waiting-reachability},
  numbers=none]
 €This algorithm is obtained from the standard Algorithm~\ref{algo:standard-reachability} by changing line~\ref{algo:standard-reachability:search-order} to€
    take and remove $(q, Z)$ minimal w.r.t. $\topopo$ from $W$
\end{lstlisting}

In the remaining of the section we will propose some simple ways of
finding a suitable $\topopo$ order.

\subsection{Topological-like ordering for a timed automaton}
\label{sec:joint:one-automaton}

It is helpful to think of the order $\topopo$ on states as some sort of
topological ordering, but we cannot really assume this since the
graphs of our automata may have loops. Given a timed automaton $\Aa$,
we find a linear order on the states of $\Aa$ in two steps:
\begin{enumerate}
\item we find a maximal subset of transitions of $\Aa$ that gives a
  graph $\Aa_{DAG}$ without cycles;
\item then we compute a topological ordering of this graph.
\end{enumerate}

Given an automaton $\Aa$, the graph $\Aa_{DAG}$ can be computed by
running a depth-first search (DFS) from the initial state of
$\Aa$. While traversing $\Aa$, we only consider the transitions that
point downwards or sideways; in other words we ignore all the
transitions that lead to a state that is on the current search
stack. At the end of the search, when all the states have been
visited, the transitions that have not been ignored form a graph
$\Aa_{DAG}$.

As an example, consider the timed automaton $\Aa$ in
Figure~\ref{fig:racing}. The transition from $q_4$ to $q_1$ is ignored
when computing $\Aa_{DAG}$ starting from $q_1$. A topological-like
ordering is computed from the resulting graph:
$q_1 \topopo q_2 \topopo q_3 \topopo q_4$. Let us see how $\topopo$
helps Algorithm~\ref{algo:standard-reachability} to explore bigger
nodes first. Starting from node $(q_1, true)$,
Algorithm~\ref{algo:standard-reachability} adds $(q_2, true)$ and
$(q_3, y>1)$ to the waiting list. Since $q_2 \topopo q_3$, the
algorithm then explores node $(q_2, true)$, hence adding node
$(q_3, true)$ to the waiting list. The small node $(q_3, y>1)$ is then
automatically erased, and the exploration proceeds from the big node
$(q_3, true)$. Observe that the exploration of the node $(q_3, y>1)$
is postponed until the second path reaches $q_3$. Upon this stage, the
zone inclusion relation will help to stop all explorations of smaller
nodes; in our case it is $(q_3,y>1)$. Thus, the algorithm performs
optimally on this example, no exploration step can be avoided.

\subsection{Topological-like ordering for networks of timed automata}
\label{sec:joint:network-automata}

Real-time systems often consist of several components that interact
with each other. In order to apply the same approach we need to find
an ordering on a set of global states of the system. For this we will
find an ordering for each component and then extend it to the whole
system without calculating the set of global states. 

We suppose that each component of a system is modelled by a timed automaton
$\Aa_i = (Q_i, {q_0}_i, F_i, X_i, Act_i, T_i)$. The system is modelled
as the product $\Aa = (Q, q_0, F, X, Act, T)$ of the components
$(\Aa_i)_{1 \le i \le k}$. The states of $\Aa$ are the tuples of
states of $\Aa_1, \dots, \Aa_k$: $Q = Q_1 \times \cdots \times Q_k$
with initial state $q_0 = \tuple{ {q_0}_1, \dots, {q_0}_k }$ and final
states $F = F_1 \times \cdots \times F_k$. Clocks and actions are
shared among the processes: $X = \bigcup_{1 \le i \le k} X_i$ and
$Act = \bigcup_{1 \le i \le k} Act_i$. Interactions are modelled by the
synchronisation of processes over the same action. There is a
transition
$(\tuple{q_1, \dots, q_n}, g, R, a, \tuple{q_1', \dots, q_n'}) \in T$
if
\begin{itemize}
\item either, there are two processes $i$ and $j$ with transitions
  $(q_i, g_i, R_i, \mathbf{a}, q_i') \in T_i$ and
  $(q_j, g_j, R_j, \mathbf{a}, q_j') \in T_j$ such that
  $g = g_i \land g_j$ and $R = R_i \cup R_j$, and $q_l' = q_l$ for
  every process $l \neq i,j$ (synchronised action)
\item or there is a process $i$ with transition
  $(q_i, g, R, a, q_i') \in T_i$ such that for every process
  $l \neq i$, $a \not \in Act_l$ and $q_l' = q_l$ (local action).
\end{itemize}

The product above allows synchronisation of $2$ processes at a time.
Our work does not rely on a specific synchronisation policy, hence
other models of interactions (broadcast communications, $n$-ary
synchronisation, etc.) could be considered as well. Notice that the
product automaton $\Aa$ is, in general, exponentially bigger than each
component $\Aa_i$.

The semantics of a network of timed automata $(\Aa_i)_{1 \le i \le k}$
is defined as the semantics of the corresponding product automaton
$\Aa$. As a result, the reachability problem for
$(\Aa_i)_{1 \le i \le k}$ reduces to the reachability problem in
$\Aa$.

In order to apply the same approach as
above, an ordering must be defined on the states of $\Aa$ which are
tuples $\vec{q} = \langle q_1, \dots, q_k \rangle$ of states of the
component automata $\Aa_i$. It would not be reasonable to compute the
product automaton $\Aa$ as its size grows exponentially with the
number of its components. 
We propose an alternative solution that consists in computing a
topological-like ordering $\topopo^i$ for each component $\Aa_i$. To
that purpose, we can apply the algorithm introduced in the previous
section. Then, the ordering of tuples of states is defined pointwise:

\begin{definition}[Joint ordering]
  For $\vec{q}, \vec{q'} \in Q_1 \times \cdots \times Q_k$, we have
  $\vec{q} \topopo \vec{q'}$ if $\vec{q}_i \topopo^i \vec{q'}_i$ for
  all $1 \le i \le k$.
\end{definition}

Thus for networks of timed automata we consider the joint ordering in
our waiting strategy.


\section{Experimental evaluation}
\label{sec:experiments}
We present and comment the experimental results that we have
performed. The results indicate that a mix of a ranking and waiting
strategies avoids mistakes in most the examples.  

We have evaluated the ranking system (Section~\ref{sec:ranking}) and
the waiting strategy (Section~\ref{sec:joint}) on classical benchmarks
from the literature\footnote{The models are available from
  \url{http://www.labri.fr/perso/herbrete/tchecker}.}:
\textsc{Critical Region (CR)}, \textsc{Csma/Cd (C)}, \textsc{Fddi
  (FD)}, \textsc{Fischer (Fi)}, \textsc{Flexray (Fl-PL)} and
\textsc{Lynch (L)}, and on the \textsc{BlowUp (B)} example in
Figure~\ref{fig:blowup}. These automata have no reachable accepting
state, hence forcing algorithms to visit the entire state-space of the
automata to prove unreachability.

Our objective is to avoid mistakes during exploration of the
state-space of timed automata. At the end of the run of the algorithm,
the set of visited nodes $P$ forms an invariant showing that accepting
nodes are unreachable. Every node that is visited by the algorithm and
that does not belong to $P$ at the end of the run is useless to prove
unreachability. This happens when the algorithm does a mistake: it
first visits a small node before reaching a bigger node later. We aim
at finding a search order that visits bigger nodes first, hence doing
as few mistakes as possible. Notice that it is not always possible to
completely avoid mistakes since the only paths to a big node may have
to visit a small node first.

We compare three algorithms in Table~\ref{table:benchmarks}:
\textsf{BFS} the standard breadth-first search
algorithm\footnote{Algorithm~\ref{algo:standard-reachability} is
  essentially the algorithm that is implemented in
  UPPAAL\cite{Behrmann:QEST:2006}.}
(i.e. Algorithm~\ref{algo:standard-reachability}), \textsf{R-BFS}
which implements a breadth-first search with priority to the highest
ranked nodes (i.e. Algorithm~\ref{algo:ranking-reachability}) and
\textsf{TW-BFS} which combines giving highest priority to true-zone
nodes and the waiting strategy. We report on the number of visited
nodes, the number of mistakes, the maximum number of stored nodes, and
the final number of stored nodes. We also mention in column ``visited
ranking'' the number of nodes that are re-visited to update the rank
of the nodes by algorithm \textsf{R-BFS}
(line~\ref{algo:ranking-reachability:update-rank} in
Algorithm~\ref{algo:ranking-reachability}). The number of visited
nodes gives a good estimate of the running time of the algorithm,
while the maximal number of stored nodes gives a precise indication of
the memory used for the set $P$.

\medskip

The ranking system gives very good results on all models except
\textsc{Csma/Cd}. It makes no mistakes on \textsc{Fischer} and
\textsc{Lynch}. This is due to the highest priority given to true-zone
nodes. Indeed, column ``visited ranking'' shows that ranks are never
updated, hence the nodes keep their initial rank. It also performs
impressively well on \textsc{BlowUp}, \textsc{Fddi} and
\textsc{Flexray}, gaining several orders of magnitude in the number of
mistakes. However, it makes more mistakes than \textsf{BFS} on
\textsc{Csma/Cd}. Indeed, the ranking system is efficient when big
nodes are reached quickly, as the example in
Figure~\ref{fig:stopRacing} shows. When the big node $(q_3, Z_3')$ is
reached, the ranking system stops the exploration of the subtree of
the small node $(q_3, Z_3)$ at $(q_4, Z_4)$. However, making the path
$q_1 \rightarrow q_2 \rightarrow q_3$ longer in the automaton in
Figure~\ref{fig:racing} leads to explore a bigger part of the subtree
of $(q_3, Z_3)$. If this path is long enough, the entire subtree of
$(q_3, Z_3)$ may be visited before $(q_3,Z_3')$ is reached. The
ranking system does not provide any help in this situation. This bad
scenario occurs in the \textsc{Csma/Cd} example.

\medskip

We have experimented the waiting strategy separately (not reported in
Table~\ref{table:benchmarks}). While the results are good on some
models (\textsc{BlowUp}, \textsc{Fddi}, \textsc{Csma/Cd}), the waiting
strategy makes a lot more mistakes than the standard \textsf{BFS} on
\textsc{Lynch} and \textsc{Flexray}. Indeed, the waiting strategy is
sensitive to the topological ordering. Consider the automaton in
Figure~\ref{fig:racing} with an extra transition $q_3 \rightarrow q_2$.
The loop on $q_2$ and $q_3$ may lead to different topological
orderings, for instance $q_1 \topopo q_2 \topopo q_3 \topopo q_4$ and
$q_1 \topopo q_3 \topopo q_2 \topopo q_4$. These two choices lead to
very different behaviours of the algorithm. Once the initial node has
been explored, the two nodes $(q_3, y>1)$ and $(q_2, true)$ are in the
waiting queue. With the first ordering, $(q_2, true)$ is selected
first and generates $(q_3, true)$ that cuts the exploration of the
smaller node $(q_3, y>1)$. However, with the second ordering
$(q_3, y>1)$ is visited first. As a result, $(q_3, true)$ is reached
too late, and the entire subtree of $(q_3, y>1)$ is explored
unnecessarily. We have investigated the robustness of the waiting
strategy w.r.t. random topological orderings for the models in
Table~\ref{table:benchmarks}. The experiments confirm that the waiting
strategy is sensitive to topological ordering. For most models, the
best results are achieved using the topological ordering that comes
from running a DFS on the automaton as suggested in
Section~\ref{sec:joint:one-automaton}.

\medskip

The two heuristics perform well on different models. This suggests to
combine their strengths.
Consider again the automaton in Figure~\ref{fig:racing} with an extra
transition $q_3 \rightarrow q_2$. As explained above, due to the cycle
on $q_2$ and $q_3$, several topological orderings are possible for the
waiting strategy. The choice of
$q_1 \topopo q_3 \topopo q_2 \topopo q_4$ leads to a bad situation
where $(q_3, y>1)$ is taken first when the two nodes $(q_3, y>1)$ and
$(q_2, true)$ are in the waiting queue. As a result, the node
$(q_3, y>1)$ is visited without waiting the bigger node $(q_3, true)$.
In such a situation, combining ranking and the waiting strategies
helps. Indeed, after $(q_3, y>1)$ has been explored, the waiting queue
contains two nodes $(q_2, true)$ and $(q_4, 1 < y \le 5)$. Since
$q_2 \topopo q_4$, the algorithm picks $(q_2, true)$, hence generating
$(q_3, true)$. As a true-zone node, $(q_3, true)$ immediately gets a
higher rank than every waiting node. Exploring $(q_3, true)$ generates
$(q_4, y\le 5)$ that cuts the exploration from the small node
$(q_4, 1<y\le 5)$.

We have tried several combinations of the two heuristics. The best one
consists in using the waiting strategy with priority to true zones.
More precisely, the resulting algorithm \textsf{TW-BFS} selects a
waiting node as follows:
\begin{itemize}
\item True-zone nodes are taken in priority,
\item If there is no true-zone node, the nodes are taken according to
  the waiting strategy, and in BFS order.
\end{itemize}

As Table~\ref{table:benchmarks} shows, \textsf{TW-BFS} makes no
mistake on all models but three. \textsc{Critical Region} has
unavoidable mistakes: big nodes that can only be reached after
visiting a smaller node. The topological ordering used for
\textsc{Fddi} is not optimal. Indeed, there exists an optimal
topological search order for which \textsf{TW-BFS} makes no mistake,
but it is not the one obtained by the algorithm presented in
Section~\ref{sec:joint:one-automaton}. Finally, the algorithm makes a
lot of mistakes on \textsc{Flexray}, but the memory usage is almost
optimal: the mistakes are quickly eliminated. This example is the only
one where applying the ranking heuristic clearly outperforms
\textsf{TW-BFS}.

We have also evaluated \textsf{TW-BFS} using randomised versions of
the models in Table~\ref{table:benchmarks}. Randomisation consists in
taking the transitions in a non-fixed order, hence increasing the
possibility of racing situations like in
Figure~\ref{fig:racing-situation}. The experiments show that the
strategies are robust to such randomisation, and the results on random
instances are very close to the ones reported in the table.

\medskip

The ranking strategy \textsf{R-BFS} requires to keep a tree structure
over the passed nodes. Using the classical left child-right sibling
encoding, the tree can be represented with only two pointers per
node. This tree is explored when the rank of a node is updated
(line~\ref{algo:ranking-reachability:update-rank} in
Algorithm~\ref{algo:ranking-reachability}). Column ``visited ranking''
in Table~\ref{table:benchmarks} shows that these explorations do not
inflict any significant overhead in terms of explored nodes, except
for \textsc{Csma/Cd} and \textsc{Critical Region} for which it has
been noticed above that algorithm \textsf{R-BFS} does not perform
well. Furthermore, exploring the tree is inexpensive since the visited
nodes, in particular the zones, have already been computed.
Both the ranking strategy and the waiting strategy require to sort the
list of waiting nodes. Our prototype implementation based on insertion
sort is slow. However, preliminary experiments show that implementing
the list of waiting nodes as a heap turns out to be very efficient.

\medskip

To summarise we can consider our findings from a practical point of
view of an implementation. The simplest to implement strategy would be
to give priority to true zones. This would already give some
improvements, but for example for \textsc{Fddi} there would be no
improvement since there are no true zones. \textsf{R-BFS} gives very
good results on \textsc{Flexray} model its implementation is more
complex than \textsf{TW-BFS} strategy is relatively easy to implement
and has very good performance on all but one model, where it is
comparable to standard \textsf{BFS}. This suggests that
\textsf{TW-BFS} could be used as a replacement for \textsf{BFS}.

\begin{table}
  \centering
  \tiny
  \tabcolsep=0.06cm
  \begin{tabular}{|l |rrrr |rrrrr |rrrr|}
    \hline
    \multirow{3}{*}{}
    & \multicolumn{4}{c|}{\textsf{BFS}}
    & \multicolumn{5}{c|}{\textsf{R-BFS}}
    & \multicolumn{4}{c|}{\textsf{TW-BFS}}
    \\
    \cline{2-14}
    & \multirow{2}{*}{visited}
    & \multirow{2}{*}{mist.}
    & \multicolumn{2}{c|}{stored} 
    & \multirow{2}{*}{visited}
    & \multirow{2}{*}{mist.}
    & \multicolumn{2}{c}{stored}
    & visited
    & \multirow{2}{*}{visited}
    & \multirow{2}{*}{mist.}
    & \multicolumn{2}{c|}{stored}
    \\

    & & & \multicolumn{1}{c}{final} & \multicolumn{1}{c|}{max}
    & & & \multicolumn{1}{c}{final} & \multicolumn{1}{c}{max} & ranking
    & & & \multicolumn{1}{c}{final} & \multicolumn{1}{c|}{max}
    \\
    \hline
    B-5
    & 63
    & 52
    & 11
    & 22

    & 16
    & 5
    & 11
    & 11
    & 13

    & 11
    & 0
    & 11
    & 11
    \\
    B-10
    & 1254
    & 1233
    & 21
    & 250

    & 31
    & 10
    & 21
    & 21
    & 28

    & 21
    & 0
    & 21
    & 21
    \\
    B-15
    & 37091
    & 37060
    & 31
    & 6125

    & 46
    & 15
    & 31
    & 31
    & 43

    & 31
    & 0
    & 31
    & 31
    \\
    & & & & & & & & & & & & & \\
    FD-8
    & 2635
    & 2294
    & 341
    & 439

    & 437
    & 96
    & 341
    & 341
    & 579

    & 349
    & 8
    & 341
    & 341
    \\
    FD-10
    & 10219
    & 9694
    & 525
    & 999

    & 684
    & 159
    & 525
    & 525
    & 1168

    & 535
    & 10
    & 525
    & 525
    \\
    FD-15
    & 320068
    & 318908
    & 1160
    & 18707

    & 1586
    & 426
    & 1160
    & 1160
    & 4543

    & 1175
    & 15
    & 1160
    & 1160
    \\
    & & & & & & & & & & & & & \\
    C-10
    & 39698
    & 5404
    & 34294
    & 48286

    & 59371
    & 25077
    & 34294
    & 52210
    & 54319

    & 34294
    & 0
    & 34294
    & 34302
    \\
    C-11
    & 98118
    & 17233
    & 80885
    & 124220

    & 153042
    & 72157
    & 80885
    & 130557
    & 160822

    & 80885
    & 0
    & 80885
    & 80894
    \\
    C-12
    & 239128
    & 50724
    & 188404
    & 311879

    & 378493
    & 190089
    & 188404
    & 320181
    & 430125

    & 188404
    & 0
    & 188404
    & 188414
    \\
    & & & & & & & & & & & & & \\
    Fi-7
    & 11951
    & 4214
    & 7737
    & 7738

    & 7737
    & 0
    & 7737
    & 7737
    & 0

    & 7737
    & 0
    & 7737
    & 7737
    \\
    Fi-8
    & 40536
    & 15456
    & 25080
    & 25082

    & 25080
    & 0
    & 25080
    & 25080
    & 0

    & 25080
    & 0
    & 25080
    & 25080
    \\
    Fi-9
    & 135485
    & 54450
    & 81035
    & 81038

    & 81035
    & 0
    & 81035
    & 81035
    & 0

    & 81035
    & 0
    & 81035
    & 81035
    \\
    & & & & & & & & & & & & & \\
    L-8
    & 45656
    & 15456
    & 30200
    & 30202

    & 30200
    & 0
    & 30200
    & 30200
    & 0

    & 30200
    & 0
    & 30200
    & 30200
    \\
    L-9
    & 147005
    & 54450
    & 92555
    & 92558

    & 92555
    & 0
    & 92555
    & 92555
    & 0

    & 92555
    & 0
    & 92555
    & 92555
    \\
    L-10
    & 473198
    & 186600
    & 286598
    & 286602

    & 286598
    & 0
    & 286598
    & 286598
    & 0

    & 286598
    & 0
    & 286598
    & 286598
    \\
    & & & & & & & & & & & & & \\
    CR-3
    & 1670
    & 447
    & 1223
    & 1223

    & 1532
    & 309
    & 1223
    & 1223
    & 1837

    & 1563
    & 340
    & 1223
    & 1223
    \\
    CR-4
    & 21180
    & 7440
    & 13740
    & 13740

    & 17694
    & 3954
    & 13740
    & 13740
    & 24295

    & 19489
    & 5749
    & 13740
    & 13740
    \\
    CR-5
    & 285094
    & 113727
    & 171367
    & 171367

    & 216957
    & 45590
    & 171367
    & 171367
    & 307010

    & 257137
    & 85770
    & 171367
    & 171367
    \\
    & & & & & & & & & & & & & \\
    Fl-PL
    & 881214
    & 228265
    & 652949
    & 652949

    & 655653
    & 2704
    & 652949
    & 652949
    & 6977

    & 12660557
    & 11997402
    & 663155
    & 684467
    \\
    \hline
  \end{tabular}
  \caption{Experimental results: \textsf{BFS} corresponds to
    Algorithm~\ref{algo:standard-reachability} with a BFS order on the
    waiting nodes, \textsf{R-BFS} implements the
    ranking system on top of the \textsf{BFS} algorithm
    (i.e. Algorithm~\ref{algo:ranking-reachability}), and
    \textsf{TW-BFS} implements the waiting strategy with a priority to 
    true-zone nodes.}
  \label{table:benchmarks}
\end{table}



\section{Conclusion}
We have analysed the phenomenon of mistakes in the zone based
reachability algorithm for timed automata. This situation occurs when
the exploration algorithm visits a node that is later removed due to a
discovery of a bigger node. It is well known that DFS exploration may
suffer from an important number of mistakes. We have exhibited
examples where BFS makes an important number of mistakes that can be
avoided.

To limit the number of mistakes in exploration we have proposed two
heuristics: \emph{ranking system} and the \emph{waiting strategy}.
The experiments on standard models show that, compared with the
standard BFS reachability algorithm the strategies using our heuristics
give not only a smaller number of visited nodes, but also a smaller
number of stored nodes. Actually, on most examples our strategies are
optimal as they do not make any mistakes. In addition, the experiments
indicate that the \textsf{TW-BFS} strategy works often as good as the
combination of both waiting and ranking strategies, while its
implementation is much simpler. Therefore, we suggest to use the
\textsf{TW-BFS} algorithm instead of standard BFS for reachability
checking.

\paragraph{Acknowledgements.}
The authors wish to thank Igor Walukiewicz for the many helpful
discussions.


\bibliographystyle{splncs03}
\bibliography{ref}

\end{document}